\documentstyle [12pt,a4]{article}

\newcommand{\be}{\begin{equation}}
\newcommand{\ee}{\end{equation}}
\newcommand{\bea}{\begin{eqnarray}}
\newcommand{\ena}{\end{eqnarray}}
\newcommand{\sect}[1]{\setcounter{equation}{0}\section{#1}}
\newcommand{\vs}[1]{\rule[- #1 mm]{0mm}{#1 mm}}
\newcommand{\hs}[1]{\hspace{#1 mm}}
\newcommand{\sm}[2]{\frac{\mbox{\footnotesize #1}\vs{-2}}
                   {\vs{-2}\mbox{\footnotesize #2}}}

\newcommand{\shalf}{\sm{1}{2}}
\newcommand{\smt}{\sm{3}{2}}
\newcommand{\smtt}{{3/2}}
\newcommand{\var}{\varphi}

\newcommand{\bz}{\bar{z}}
\newcommand{\bh}{\bar{h}}
\newcommand{\cd}{\mbox{$\cal{D}$}}

\newcommand{\cg}{\mbox{$\cal{G}$}}
\newcommand{\cs}{\mbox{$\cal{S}$}}

\newcommand{\ct}{\mbox{$\cal{T}$}}

\newcommand{\cc}{\mbox{$\cal{C}$}}
\newcommand{\cj}{\mbox{$\cal{J}$}}
\newcommand{\calh}{\mbox{$\cal{H}$}}
\newcommand{\prt}{\partial}
\newcommand{\al}{\alpha}

\newcommand{\mb}[1]{\hs{7}\mbox{#1}\hs{7}}
\newcommand{\NP}[1]{Nucl.\ Phys.\ {\bf #1}}
\newcommand{\PL}[1]{Phys.\ Lett.\ {\bf #1}}

\newcommand{\CMP}[1]{Comm.\ Math.\ Phys.\ {\bf #1}}

\begin{document}
\renewcommand{\thefootnote}{\fnsymbol{footnote}}
\newpage
\setcounter{page}{0}

\vs{15}

\begin{center}

{\LARGE {\bf General Properties of Classical $W$ Algebras \footnote{Plenary
talk presented by P. SORBA at the $XXII^{th}$ International Conference on
Differential Geometric Methods in Theoretical Physics. Ixtapa Mexico, September
1993.}
\footnote{Supported in part by EEC Science Contract SC10000221-C}}}\\[1cm]

{\large F. Delduc${}^1$, L. Frappat${}^2$, E. Ragoucy${}^2$,\\[.3cm]
and P. Sorba${}^{1,2}$}\\[.5cm]
{\em Laboratoire de Physique Th\'eorique}
{\small E}N{\large S}{\Large L}{\large A}P{\small P}
\footnote{URA 14-36 du CNRS, associ\'ee \`a l'E.N.S. de Lyon, et au L.A.P.P.
(IN2P3-CNRS) d'Annecy-le-Vieux \\
${}^1$ Goupe de Lyon, ENS Lyon,
46 All\'ee d'Italie, F-69364 LYON CEDEX 07, France\\
${}^2$ Groupe d'Annecy, LAPP, BP 110,
F - 74941 Annecy-le-Vieux Cedex, France}
\end{center}
\vfill

\centerline{\bf Abstract}

\indent

After some definitions, we review in the first part of
this talk the construction and classification of classical $W$ (super)algebras
symmetries of Toda theories. The second part deals with more recently obtained
properties. At first, we show that chains of $W$ algebras can be obtained by
imposing constraints on some $W$ generators: we call secondary reduction such a
gauge procedure on $W$ algebras. Then we emphasize the role of the Kac-Moody
part, when it exists, in a $W$ (super) algebra. Factorizing out this spin 1
subalgebra gives rise to a new $W$ structure which we interpret either as a
rational finitely generated $W$ algebra, or as a polynomial non linear
$W_\infty$ realization.
\vfill

\rightline{{\small E}N{\large S}{\Large L}{\large A}P{\small P}-AL-449/93}
\rightline{November 1993}



\newpage

\renewcommand{\thefootnote}{\arabic{footnote}}
\setcounter{footnote}{0}
\newpage

\sect{Introduction \label{sect:1}}

\indent

$W$ algebras constitute today a rather broad subject: on the one hand
they play a role in different parts of 2 dimensional Conformal Field Theories
(CFT), on the other hand much has still to be done for a complete knowledge
of these algebras and their algebraic properties. First it was thought that
they can be used to facilitate the analysis of rational CFT (i.e. theories in
which the main parameters, namely central charge $c$ and conformal dimensions
$h_i$ are all rational numbers): this extra symmetry, bigger than the conformal
one, could help to characterize degeneracies, and to classify in a
simpler way the
physical states. After that it was realized that
they show up in several places. We currently
talk nowadays about $W$ gravity. $W$ algebras
appear in the quantum Hall effect,
black holes models, in lattice models of statistical mechanics at
criticality, and
in Toda models\cite{sav} as symmmetry algebras \cite{ORaf}.

\indent

After some definitions (Section \ref{sect:2}), we will
concentrate on classical $W$ algebras and
superalgebras which are finitely generated -we generically
denote them $W_n$-. Two
remarkable facts can then be mentioned (Section \ref{sect:3}):

-i) The constants of motion of a Toda theory form a $W_n$ algebra,
and such a Toda
theory can be seen as a gauged WZW model, on which constraints have been
imposed \cite{ORaf}.

-ii) As a consequence, one can explicitly construct such $W_n$
algebras, and
give a group theoretical classification of them \cite{nous}.

Two comments:

- this classification is based on the $Sl(2)$ embeddings in a simple Lie
(super)algebra $\cg$ and on the $OSp(1|2)$ embeddings
in a simple superalgebra ${\cs}
\cg$. We will try to insist on the property of
$Sl(2)$ to be intimately linked to a
$W_n$ algebra from its definition: this is important for our construction,
but also allows to think that the classification of $W_n$ algebras
symmetries of Toda models hereafter given is "not far"
from exhausting the set of $W_n$ algebras.

- there are two main types of $W_n$ algebras: those that we will call the
Abelian ones because they are related to Abelian Toda models:
for example, if the underlying group of the Toda model is $Sl(n)$, one
gets the algebra generated by $W_2, W_3, ...W_n$.

There is a second type of $W_n$ algebra, less well-known:
they are associated to non
Abelian Toda models\cite{sav}, and we call them non
Abelian $W_n$ algebras, and we will
come back to this class of algebras.

The above classification can be simplified using two interesting features,
directly suggested by properties of  simple Lie algebras and
superalgebras, namely:

- deduction of $W_n$ algebras related to non simply laced algebras $B_n,
C_n...$ from $W_n$ algebras related to $A_n$ series by "foldings"
\cite{fold} analogous to
the folding technics which produce
$B_n, C_n...$ algebras from $A_n$ ones (Section
\ref{sect:4}).

- existence of chains of $W_n$ algebras mimicking chains of embeddings of
subalgebras in a simple Lie Algebra \cite{prep}.
Imposing constraints, when possible,
on a the $W$ algebra itself, one can reduce $W$ into
another algebra $W$: we will call this technics a secondary reduction (Section
\ref{sect:5}).

Finally coming back to the non Abelian $W_n$ algebras, one can remark that most
of them contain a Kac Moody part. Such a Kac Moody subalgebra should
play a
particular role. In particular, we will see that factorizing out this "spin
one" part in the $W_n$ algebra gives rise to an algebra which can be seen
either as an $W_\infty$ algebra, that is an infinitely generated $W$ algebra,
or as a finitely generated $W$ algebra but of a new type; we will call it
"rational" $W_n$ algebras \cite{Wrat}.
This problem as well as its supersymmetric
generalisation is the subject of Section  \ref{sect:6}. which ends up by a
comparative study of the factorizations of spin $1/2$ fermions and spin 1
bosons in a $W$ algebra.

We have chosen to illustrate each property which is introduced on an example
instead of presenting general proofs. We hope that this approach will make the
reading as easy for the non experts as for those familiar with $W$ algebras,
these last ones being invited to directly go to the three last sections.

\sect{Definitions \label{sect:2}}

\indent

We know from $d=2$ CFT that the stress energy tensor has a short-distance
O.P.E. of the form, with  $z,w$ complex variables:

\be
T(z).T(w) = \frac{2T(w)}{(z-w)^2} + \frac{\prt T(w)}{(z-w)^2} +
\frac{c/2}{(z-w)^4} +\dots
\ee

Expressing $T(z)$ into Laurent modes
\be
T(z) = \sum_{m \in Z} z^{-m-2} L_m \ \ \ \ \ \ L_m = \oint \frac{dz}{2i\pi z}
 \ z^{m+2} T(z)
\ee
the integral being understood around the origin clockwise, we have the C.R. of
the Virasoro algebra:
\be
[L_m, L_n] = (m-n) L_{m+n} + \sm{c}{12} m(m^2-1)  \delta_{m+n,0}
\ee

Note that $\{ L_{+1}, L_{-1}, L_0\}$ generate an $Sl(2,R)$ algebra, while $c$
is the central charge.

In a CFT, primary fields are those which transform as tensors of weight $(h,
\bar{h})$ under conformal transformations:
\[
z \rightarrow w(z), \ \ \bar{z} \rightarrow \bar{w} (\bar{z})
\]
\be
\phi'_{h,\bh} (z,\bz) = \phi_{h,\bh} \left( w(z), \bar{w}(\bz) \right)
\left( \frac{dw}{dz} \right)^h \left( \frac{d\bar{w}}{d \bz} \right)^{\bh}
\ee
$T(z)$ being the generator of local scale transformations, one gets the
O.P.E., after restricting to the $z$-part:
\be
T(z) . \phi_h (w) = \frac{h \phi_h (w)}{(z-w)^2} + \frac{\prt \phi_h
(w)}{(z-w)} + ... \label{eq.2.5}
\ee
$h$ is called the conformal spin of the primary field $\phi_h(z)$. One can
deduce from eq. (\ref{eq.2.5}) the CR:
\be
\left[ L_m, \phi_h(z) \right] = (m+1) hz^m \phi_h(z) + z^{m+1} \prt \phi_h(z)
\ee

Now let us add to the Virasoro algebra some primary fields. With some
precautions, we can obtain a $W$ algebra.

As an example, let us consider the
$N=1$ superconformal algebra: it is made from the (conformal spin 2) stress
energy tensor $T(z)$ and a conformal spin $3/2$ fermionic field $G(z)$.
Developing $T(z)$ and $G(z)$ in Laurent modes:
\be
G(z) = \sum z^{-3/2-r} \ G_r
\ee
with $r \in {\bf Z}$ or $r \in {\bf Z} + \frac{1}{2}$
following we are in the Ramond
or Neveu-Schwarz sector, we get the (anti) C.R.:
\bea
\left[ L_m, G_r \right] &=& (\shalf m-r) G_{m+r} \nonumber \\
\{ G_r, G_s \} &=& 2 L_{r+s} + \sm{c}{3} (r^2 -1/4) \delta_{r+s,0}
\ena

We have a $W$ (super)algebra. It is specially simple since it closes linearly
on the generators $L_m$ and $G_r$. Let us add two remarks which will be
relevant for the future.

First $\left\{ L_{+1}, L_{-1}, L_0, G_{+1/2}, G_{-1/2} \right\}$ generate the
$OSp(1|2)$ superalgebra, that is the "supersymmetric" $Sl(2)$ extension. In the
following $OSp(1|2)$ will play for $W_n$ superalgebras the role of $Sl(2,R)$
for $W_n$ algebras.

Secondly $\{ G_{\pm{1/2}} \}$ constitutes a spin $1/2$ representation of the
algebra $\{ L_{\pm1}, L_0 \}$. More generally \cite{7} if $W_h(z)$ is a $h$
primary field under $T(z)$ the modes $W_n$ with $-h+1 \leq n \leq h-1$ will
form a spin $(h-1)$ representation of $\{L_{\pm1}, L_0 \}$.

The above definitions and properties stand for the above OPE to be
radially ordered. We will
relax this last feature in the following and restrict ourselves to the
classical case.

Then a classical finitely generated $W_n$ algebra will be defined as a Lie
algebra with a Poisson bracket $\{ , \}_{P.B.}$, and a set of
generators involving a stress-energy tensor $T$
as well as a finite number of primary fields
$W_{h_i}(i=1,...n-1)$ under $T$ satisfying:

\[
\{ T(z) , T(w) \}_{P.B.} = -2T(w) \delta'(z-w) + \prt T(w)
\delta(z-w) +
\]
\be
+\sm{c}{2} \delta''' (z-w)
\ee
\be
\{ T(z) , W_{h_i} (w) \}_{P.B.} = -h_i W_{h_i}(w) \delta'(z-w) + \prt
W_{h_i} (w) \delta(z-w)
\ee
and
\be
\{ W_{h_i}(z), W_{h_j}(w) \} = \sum_\alpha P_{i,j;\alpha} (w)
\delta^{(\alpha)} (z-w)
\ee
where $P_{i,j;\alpha}(w)$ are polynomials in the primary
fields $W_{h_i}, T$ and their derivatives.

Let us remark that the property of a primary field $W_h$ of conformal spin $h$
to be connected to the representation $D_{h-1}$ of the $Sl(2,R)$ algebra
$\{L_\pm , L_0\}$ limitates through the tensorial product $D_{h_i-1} \times
D_{h_j-1}$ the allowed conformal spin of the $P_{i,j;\alpha}$ polynomials.

\sect{From a WZW model to a Toda theory \label{sect:3}}
\subsection{The method}

\indent

It has been elegantly shown that, starting from a WZW model, the action of
which is $S(g)$ and the fields $g(x)$ belong to the group $G$, and imposing
some of the components of the conserved currents to
be constant or zero leads to
a Toda model \cite{ORaf}.

Let us denote $S_{WZW}(g)$ the action of the WZW model based on a real
connected Lie group $G$,
and $g \in G$. Then from the Kac-Moody invariance $G_1 \times G_2$ with $G_1
\cong G_2 \cong G$ of the model
\be
g(x) \rightarrow g_1 (x^-) g(x) g_2(x^+)
\ee
with $x=(x^+,x^-)$ denoting the two-dimensional variable, we get the currents:
\be
J_+ = g^{-1} \prt_+ \ g \ \ \mbox{and} \ \ J_- = \prt_- g g^{-1}
\ee
which, due to the equations of motion, are conserved:
\be
\prt_{\pm} J_{\mp} =0
\label{eq:2.3}
\ee

In order to perform the gauge theory approach which will be relevant, we need
$G$ to be non compact: let us consider as an example the $Sl(n,R)$ group. We
decompose its Lie algebra $\cg$ as follows:
\be
{\cg} = {\cg}_- \oplus {\calh} \oplus {\cg}_+
\ee
where $\cg_+(\cg_-)$ is the subalgebra of positive (negative) root generators
and ${\calh}$ the Cartan part, i.e.:
\be
\left(
\begin{array}{cccc}
* & & & \\
 & \ddots & \cg_+ & \\
 & \cg_- & \ddots & \\
& & & *
\end{array}
\right)
\label{eq:3.5}
\ee
Note that the generators $E_{\al_i}(i=1 ... n-1)$ associated to the (positive)
simple roots are in the positions $E_{12}, E_{23}, ... E_{n-1,n}$ in the above
matrix, while $E_{-\al_i}$ occupy the position $E_{21},...,E_{n,n-1}$ ($E_{ij}
$ being the $n \times n$ matrix with 1 in position $(i,j)$ only).

The basic idea is to impose constraints on some components of these $J_\pm$
currents. Let us impose the restriction of $J_-$ to its $\cg_-$ components
to be:
\be
\left. J_{-}\right|_{ \cg_-} = M_-  = \sum^{n-1}_{i=1} \mu_i E_{-\al_i}
\ \ \ \ \
\left. J_{+}\right|_{\cg_+} = \sum^n_{i=1} \nu_i E_{\al_i}
\label{eq:2.6}
\ee
with $\mu_i$ and $\nu_i$ real positive constants.

\indent

Such constraints can be obtained as a part of the equations of motion of a new
model resulting from a Lagrange multiplier treatment on the WZW action. More
precisely, it is a gauge theoretical approach involving as gauge group  the
(non
compact) part $G_+$ in $G_1$ and $G_-$ in $G_2$, associated to the Lie $\cg$
subalgebra $\cg_+$ and $\cg_-$ respectively with elements $g_+(x) \in G_+$
and $g_-(x) \in G_-$ which will lead to the Euler equations (\ref{eq:2.3}) and
(\ref{eq:2.6}). The use of the local Gauss decomposition
\be
g= g_+ \cdot h \cdot g_-
\ee
with
\be
h(x) = \exp \sum^r_{i=1} \phi_i(x) H_i
\ee
provides in the Euler equations the differential equations of the Toda theory
based on the group $G$, the $\phi_i$'s being the corresponding fields.
\be
\prt_+ \prt_- \phi_i = \mu_i \nu_i\ \exp \sum_j K_{ij} \phi_j
\ee
where $K_{ij}$ is the Cartan matrix associated to the Lie algebra $\cg$ of $G$.

Two remarks can be made at this point.

i) The above G Toda theory involves $r=$ rank $\cg$ fields in one-to-one
correspondence with the Cartan part $\calh$ of $\cg$, and
it is usually called the
"Abelian" Toda theory on $\cg$.

ii) The above construction actually involves the principal $Sl(2)$ subalgebra
of $\cg$ with generators:
\be
H=\sum^r_{i,j=1} K^{ij} H_j \ \ \ \
E_-=\sum^r_{i=1} E_{- \al_i} \ \ \ \ \ \
E_+ = \sum^r_{i,j=1} K^{ij} E_{\al_i}
\ee
(note that a rescaling in Eq.(\ref{eq:2.6}) allows to take all the
$\mu_i=1$; $K^{ij}$ is the inverse Cartan matrix).

Moreover the currents $J_-$ (resp. $J_+)$ are not invariant under the gauge
transformations generated by the constraints (\ref{eq:2.6}). Focussing
on $J_-$, these transformations read:
\be
J_-(x_-) \rightarrow J_-^g(x_-) =  g_+ (x_-) J_-(x_-)g_+(x_-)^{-1} + \prt_-
g_+(x_-) \cdot g_+ (x_-)^{-1}
\ee
where $g_+(x_-)\in G_+$. This will allow to bring the currents to
the gauge-fixed form:
\be
J^g = M_- + \sum_{j \geq 0} W_{j+1}(J) M_j
\ee
where the $W_{j+1}$ are polynomials in the currents $J_-$ and their derivatives
$\prt_-^n J_-$. In the so-called "Drinfeld-Sokolov highest weight gauge"
each generator $M_j$ is the highest weight in the
$Sl(2)_{ppal}$ representation ${\cg}_j$ space obtained by reducing with respect
to $Sl(2)_{ppal}$ the Lie algebra $\cg$: considered as a vector space,
$\cg$ writes
\be
{\cg} = \oplus^k_{j=1} D_j
\label{eq:3.13}
\ee
with $D_j$ of dimension $(2j+1)$. The Poisson brakets among the $W_j$'s can be
obtained from the Poisson-Lie algebra satisfied by the current components:
\be
\{ J_-^a (x_-), J_-^b (x_-') \}_{PB} = i f^{ab}_c J_-^c ( x'_-) \delta
(x_--x'_-) + k \delta^{ab} \delta ' (x_- - x'_-)
\label{eq:2.10}
\ee
where $f^{ab}_c$ are the structure constants for a given basis of $\cg$.

Then each $W_{j+1}$ is associated to a $D_j$ and its conformal spin is $(j+1)$
with respect to the stress energy tensor itself relative to the $D_1$
representation spanned by the generators of $Sl(2)_{ppal}$:
\be
T=T_0 + tr H. \prt J
\ee
with
\be
T_0 = \sm{1}{2k} tr (J.J).
\ee

Note also that each $W_{j+1}$ can always be seen
as a primary field with respect to
$T$, after adjunction of an extra term in the $J's$ and derivatives.

Before going to examples, let us remark that, in this approach, a classical
$W$-algebra is a subalgebra of the enveloping algebra of
(\ref{eq:2.10}), itself symmetry of a WZW model: the constraints
reduce the symmetry in such a way that only some polynomials in the $J^a$'s
and their derivatives generate the residual symmetry.

\subsection{Examples}

\indent

Let us take for $\cg$ the $Sl(3)$ algebra.

The Abelian Toda theory is obtained by imposing on the $J$
currents the constraints:
\be
J_-=
\left[
\begin{array}{ccc}
\varphi_1 & \varphi_3 & \varphi_4 \\
1 & \varphi_2 & \varphi_5 \\
0 & 1 & - \varphi_1 -\varphi_2
\end{array}
\right] \ \
\begin{array}{c}
\mbox{leading by the} \\
\mbox{gauge action of}\\
g_+(x_-) \in G_+ \ \mbox{to}
\end{array}
\ \
J_-^g=
\left[
\begin{array}{ccc}
0 & T & W_3 \\
1 & 0 & T \\
0 & 1 & 0
\end{array}
\right]
\label{eq:2.13}
\ee
Involving $Sl(2)_{ppal}$ generated by:
\be
E_-=
\left(
\begin{array}{ccc}
0 & 0 & 0\\
1 & 0 & 0 \\
0 & 1 & 0
\end{array}
\right) \ \
E_+ = \left(
\begin{array}{ccc}
0 & 1 & 0 \\
0 & 0 & 1\\
0 & 0 & 0
\end{array}
\right)
\ \
H =
\left(
\begin{array}{ccc}
1 & 0 &0 \\
0 & 0 & 0 \\
0 & 0 & -1
\end{array}
\right)
\ee
$\cg$ decomposes under the (adjoint) action of $Sl(2)_{ppal}$ as:
\be
{\cg}/_{Sl(2)} = D_1 \oplus D_2
\ee
to which are associated resp. with
the spin 2 and 3 quantities $T$ and $W_3$
generating the well known Zamolodchikov \cite{8} $\{T, W_3 \}$ algebra.

But still with $Sl(3)$ there exists another kind of constraints which allows
for a
similar treatment of the WZW model. It reads
\be
J_-=
\left[
\begin{array}{ccc}
\var_1 & \var_3 & \var_4 \\
1 & \var_2 & \var_5 \\
0 & \var_6 & -\var_1 -\var_2
\end{array}
\right]
\ee

Now the $Sl(2)$ subalgebra which is involved is the following:
\be
E_{-\al_1}=
\left(
\begin{array}{ccc}
0 & 0 & 0\\
1 & 0 & 0 \\
0 & 0 & 0
\end{array}
\right) \ \
E_{+\al_1} = \left(
\begin{array}{ccc}
0 & 1 & 0 \\
0 & 0 & 0\\
0 & 0 & 0
\end{array}
\right)
\ \
H =
\left(
\begin{array}{ccc}
1/2 & 0 &0 \\
0 & -1/2 & 0 \\
0 & 0 & 0
\end{array}
\right)
\label{eq:2.17}
\ee
with respect to this $Sl(2), \ \cg$ decomposes as:
\be
{\cg} = D_1 \oplus D_{1/2} \oplus D_{1/2} \oplus D_0
\ee
and the gauge invariant matrix current takes the form:
\be
J_-^g =
\left[
\begin{array}{ccc}
W_1 & W_2 & W^+_{3/2} \\
1 & W_1 & 0 \\
0 & W^-_{3/2} & -2W_1
\end{array}
\right]
\label{eq:2.19}
\ee

The algebra $\{W_2, W_{3/2}^+, W^-_{3/2}, W_1 \}$ is usually called the
classical Bershadsky algebra \cite{bersh}.
It is the symmetry algebra of the "non
Abelian" Toda model constructed from the $Sl(2)$ algebra defined in
(\ref{eq:2.17}).

\indent

There are only two different $Sl(2)$ subalgebras in $Sl(3)$; therefore we have
exhausted the different Toda models and the associated $W$-algebras relative to
$Sl(3)$. More generally, starting from a simple algebra $\cg$, each admissible
choice of $J$ components which can be set to constant (i.e. first class
constraints in Dirac terminology) will correspond to an $Sl(2)$ in $\cg$ and
vice-versa. Then to determine all the different $W$-algebras symmetries of Toda
theories associated to $\cg$, one has first to consider all the different
$Sl(2)$ in $\cg$. (This mathematical problem has been solved by Dynkin). In
each
case, the decomposition of $\cg$ with respect to $Sl(2)$ representations will
provide the conformal spin of the associated $W$ algebra \cite{nous}.

Supersymmetric Toda theories can also be considered.
A supersymmetric
treatment of the WZW models, based on simple superalgebras $\cs \cg$
has to be done, constraints being written in a superspace formulation
\cite{supertoda}.
Then $Sl(2)$ is
replaced by its supersymmetric extension $OSp(1|2)$. The classification of
$OSp(1|2)$ subsuperalgebras in simple  superalgebras followed by the reduction
for each $\cs \cg$ of its adjoint representation with respect to each
$OSp(1|2)$
subpart provide the conformal superspin content of the $W$ superalgebras
symmetries of Super Toda theories \cite{nous}.

{}From such a classification, general properties of the $W$ (super)algebras,
allowing a simplified and synthetic overview, can be deduced: this will be the
object of the two next sections.

\sect{Folding the $W$ (super)algebras \label{sect:4}}

\indent

Using the properties of a non simply laced simple algebra to appear as a
subalgebra  of $Sl(n)$ after a suitable identification of $Sl(n)$ simple roots,
one can obtain $W$ algebras  related to B-C-D series from $W$ algebras related
to unitary ones \cite{fold}.
Let us give an example, based again on the $Sl(3)$ group. Its
Dynkin diagram (DD) is :
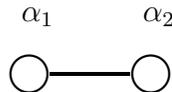
\be
\begin{picture}(100,60)
\thicklines
\put(15,30){\circle{14}} \put(12,50){\rm{$\al_1$}}
\put(23,30){\line( 1, 0){30}}
\put(61,30){\circle{14}} \put(58,50){\rm{$\al_2$}}
\end{picture}
\ee
$\al_1$ and $\al_2$ representing the simple roots, to which are associated the
generators $E_{\al_1}$ and $E_{\al_2}$. It is known that the transformation
$\tau$ such
that: $\tau(\al_i) = \al_j \ \ i \neq j=1,2$ which is a symmetry of DD can be
lifted up to an (outer) automorphism on the Lie algebra of $Sl(3)$ by defining:
\be
\hat{\tau} \left( E_{\pm\al_i} \right) = E_{\pm\tau(\al_i)} \ \ i=1,2
\ee
with
\be
\hat{\tau} [E_{\al_i}, E_{-\al_i} ] = \tau(\al_i) H
\ee

The $Sl(3)$ subalgebra $\cg$ invariant under $\hat{\tau}$ is then generated
from:
\be
E_{\pm\al_1} + E_{\pm\al_2}
\ee
That is, by "folding" the root $\al_1$ onto $\al_2$, $Sl(3)$ reduces to the Lie
algebra $\cg^F$ of the (non compact) 3 dimensional orthogonal group:
\be
\begin{picture}(160,60)
\thicklines
\put(15,30){\circle{14}}\put(12,50){\rm{$\al_1$}}\put(10,10){\rm{$E_{\al_1}$}}
\put(23,30){\line( 1, 0){ 30}}
\put(61,30){\circle{14}}\put(58,50){\rm{$\al_2$}}\put(56,10){\rm{$E_{\al_2}$}}
\put(90,30){$\longrightarrow$}
\put(150,30){\circle{14}}\put(135,50){\rm{$\al_1+\al_2$}} \put(130,10)
{\rm{$E_{\al_1} + E_{\al_2}$}}
\end{picture}
\ee
On the $3 \times 3$ matrix representation, where $E_{\al_1}$ is identified with
$E_{12}$ and $E_{\al_2}$ with $E_{23}$, it will result that from the $\cg$
matrices $M=m^{ij} E_{ij}$, $m^{ij}$ being real numbers satisfying the
traceless condition $\sum^3_{i=1} m^{ii} =0$, one obtains a representation of
$\cg^F$ by imposing the conditions:
\be
m^{ij} = (-1)^{i+j+1} m^{4-j,4-i}
\label{eq:2.22}
\ee

Identifying in the Abelian Toda theory on $Sl(3)$ the $J^a$ current
components as in (\ref{eq:2.22}), it is not a surprise to get, by Hamiltonian
reduction:
\be
J^g_{Sl(3)} =
\left(
\begin{array}{ccc}
0 & T & W_3 \\
1 & 0 & T \\
0 & 1 & 0
\end{array}
\right)
\Rightarrow J^g_{SO(3)} =
\left(
\begin{array}{ccc}
0 & T' & 0 \\
1 & 0& T'\\
0&1&0
\end{array}
\right)
\ee
as can be expected in a rank 1 algebra.

Of course, this simple example can be generalized, the foldings of $A_{2n-1} =
Sl(2n)$ and $A_{2n} = Sl(2n+1)$ providing the symplectic $C_n =Sp(2n)$ and
$B_n=SO(2n+1)$ algebras respectively. If one notes that $SO(2n)$ can be
obtained
from $SO(2n+1)$ by a regular embedding, one realizes that the $W$ algebras
associated to the $A_n$ series can be "folded" into the $W$ algebras relative
to
the other infinite series (note also that for the exceptional cases, the $G_2$
ones can be deduced from $D_4 \equiv SO(8)$ and $F_4$ $W$-algebras from the
$E_6$ ones). The same procedure can be applied to superalgebras
(see \cite{fold}).

An useful consequence of this technics is to get identities between structure
constants of $W$-algebras relative to different simple algebras:
denoting by
$C^k_{ij}$ the general structure constant of the "fusion rule":
\be
[W_i] \cdot [W_j] = \delta_{ij} \sm{c}{2} [I] + C_{ij}^k ({\cg}) [W_h]
\ee

We have as examples, in the Abelian case:
\bea
C^k_{ij}(D_n) &=& C^k_{ij}(A_{2n}) \ \ \ \ \ \ \ i,j,k \neq n\\
C^k_{ij}(C_n) &=& C^k_{ij}(A_{2n-1}) \ \ \ \ C^k_{ij} (B_n)= C^k_{ij}
(A_{2n}),
\ena
such relations being sometimes precious, due to the difficulty to obtain
explicit commutation relations.

\sect{Secondary reductions \label{sect:5}}

\indent

Let us consider again ${\cg} = SL(3)$ and the two $W$-algebras which can be
constructed, via Toda theories, from such an underlying simple
algebra; they
are the Zamolodchikov algebra $\{ T,W_3 \}$ and the Bershadsky algebra
generated by $\{ W_2, W^+_\smtt , W^-_\smtt, W_1 \}$. The corresponding $J^g$
matrices read (see Eq. (\ref{eq:2.13}) and (\ref{eq:2.19})):
\be
J^g_{\mbox{\scriptsize{Abel}}} =
\left(
\begin{array}{ccc}
0 & T & W_3 \\
1 & 0 & T \\
0 & 1 & 0
\end{array}
\right)
\ \ \ \ \ \
J^g_{\mbox{\scriptsize{Non Abel}}}=
\left(
\begin{array}{ccc}
W_1  & W_2 & W^+_\smtt \\
1 & W_1 & 0 \\
0 & W^-_\smtt & -2W_1
\end{array}
\right)
\ee
One remarks that the constraints imposed in the Non Abelian case
\be
\{ tr J_-\cdot E_{-\al_1} =1 \ \ ; \  \
tr J_-\cdot E_{-(\al_1 + \al_2)} = 0 \}
\ee
form a subset of the constraints corresponding to the Abelian case:
\be
\left\{ tr J_- \cdot E_{-\al_1} = tr J_- \cdot E_{-\al_2} =1 \ ; \ tr J_- \cdot
E_{-(\al_1 + \al_2)} =0 \right\}
\ee

It is time to give explicitly the P.B. of the Classical Bershadsky algebra:
let us, for convenience, make a little change in the notations and denote $W_1$
by $J$ and $W_2+ \frac{1}{3c}J\cdot J$ by $T$.
\bea
\{ J(z), J(w) \} &=& - \smt c\delta ' (z-w) \nonumber \\
\{ J(z), W^{\pm}_{\smtt} (w) &=& \pm \smt W^\pm_\smtt \delta (z-w) \nonumber \\
\{ T(z), W^{\pm}_{\smtt} (w) \} &=& - \smt W^\pm_\smtt (w) \delta '  (z-w) +
\prt W_\pm(w) \delta (z-w) \nonumber \\
\{ T(z), J(w) \} &=& - J(w) \delta ' (z-w) + \prt J (w) \delta (z-w) \nonumber
\\
\{ T(z), T(w) \} &=& -2T(w) \delta ' (z-w) + \prt T(w) \delta (z-w) + \sm{c}{2}
\delta '''(z-w) \nonumber \\
\{W^+_\smtt (z), W^-_\smtt (w) \} &=& 2J(w) \delta '(z-w) - c \delta''(z-w)+
\nonumber \\
&& +(T - \sm{4}{3c} J^2 - \prt J ) (w) \delta (z-w) \nonumber \\
\{W^\pm_\smtt (z), W^\pm_\smtt (w) \} &=& 0
\label{eq:4.4}
\ena

The last relation, which expresses the nilpotency of $W^-_{\smtt}$ (and
$W^+_\smtt$), allows to consider the constraint
\be
W^-_\smtt =1 \label{eq.smtt}
\ee
as a gauge constraint (first class constraint).

With the help of $J(z)$, it is possible to redefine the energy momentum tensor
$T$ in such a way that the constraint becomes conformally invariant, that is,
shifting $T$ into
\be
\hat{T} = T - \prt J_+
\label{eq:4.6}
\ee
$W^-_\smtt$ behaves as a spin 0 field:
\bea
\{ \hat{T}(z), W^-_\smtt (w) \} &=& \prt W^-_\smtt (w) \delta(z-w) \nonumber \\
\  & \simeq & 0 \ \ \ \mb{using Eq.(\ref{eq.smtt})}.
\ena

Then one can look at the reduced $W$ algebra obtained by constructing the
polynomials invariant under the gauge transformations associated to
$W^-_\smtt$.
Therefore, let us consider the finite gauge transformations on the currents:
\bea
X(w) \rightarrow \hat{X}(w) &=& X(w) + \int dz \ \al(z) \{ W^-_\smtt (z),
X(w) \} \nonumber \\
&& + \sm{1}{2!} \int dz \ dz' \al(z) \al(z') \left\{ W^-_\smtt (z), \{
W^-_\smtt
(z') , X(w) \} \right\} \nonumber \\
&& + ... \label{eq:5.8}
\ena
where $X=J,T,W^+_\smtt$, the constraint (\ref{eq.smtt}) being used on the
r.h.s. of the P.B., following Dirac prescriptions on constraints ("weak
equations"). Then the $J$ current transforms as:
\be
\hat{J}(w) = J(w) + \int dz \ \al(z) \{ W^-_\smtt (z), W_1 (w) \} +0
\ee
since
\be
\{ W^-_\smtt(z), J(w) \} \simeq \left( \shalf \delta(z-w) \right)
\ee
that is:
\be
\hat{J}(w) = J(w) + \smt \al(w)
\ee

Then, it is clear that a global gauge fixing is given by
\be
\hat{J}(w) =0
\ee
that is, by taking:
\be
\al = - \sm{2}{3} J
\ee

It follows for $T$:
\bea
T(w) \rightarrow \hat{T} (w) &=& T(w) - \smt \int dz \cdot \al(z) \cdot \delta
' (z-w) +0 \nonumber \\
&=& T(w) + \smt \prt \al \nonumber \\
&=&T - \prt J
\ena
as expected from Eq.(\ref{eq:4.6}) !

In the same way:
\be
W^+_\smtt \rightarrow \hat{W}_3 = W^+_\smtt + \sm{2}{3} J \cdot T + \sm{2}{3} J
\cdot \prt J - \sm{8}{27c} J^3 - \sm{2c}{3} \prt^2 J
\ee
the notation $\hat{W}_3$ being justified by the property of $\hat{W}_3$ to
behave as a spin 3 field under $\hat{T}$.

At this point, it is not a surprise to realize that the $\hat{T}$ and
$\hat{W_3}$ quantities generate a (algebra isomorphic to) Zamolodchikov
algebra.

The above illustrated method with $W$ algebras based on ${\cg}=SL(3)$ can be
applied to any simple algebra $\cg$ up to some obvious technical difficulties.
Starting from the weakest constraints and adding new ones on a $W$ algebra
relative to some Lie algebra $\cg$, one can then obtain chains of
$W$ algebras, the "smallest" one being relative to the Abelian Toda case
(highest number of constraints). As could be expected by Lie algebra experts,
there also exist cases with $\cg$ non simply laced, i.e. $B_n$ or $C_n$, for
which such a secondary reduction towards the Abelian case cannot be obtained.
Finally, in the same way one gets Toda equations by gauging $WZW$ models, a
gauging of the Toda action in which a (Non Abelian) $W$ algebra stands as the
current algebra of the theory could be performed, leading to a new (more
constrained) Toda action. Such an approach for a generalized gauge Toda field
theory, as well as a more complete discussion on secondary reductions will soon
be available \cite{prep}.

\sect{Rational $W$ algebras \label{sect:6}}
\subsection{Commutant of the spin 1 part}

\indent

Now let us turn our attention to the particular role of the spin one part, when
it is present, in a $W$ algebra. One can easily check, by dimensional
arguments, that these fields generate a Kac-Moody algebra $W_1$.
Moreover the set of $W$ generators decomposes into irreducible representations
under the adjoint action of this Kac Moody algebra. Let us study what happens
when factorizing out the spin one part in a $W$ algebra,
that is by computing the
commutant in $W$ of the $W_1$ Kac-Moody subalgebra \cite{Wrat}.

Most of $W$ algebras associated to Non Abelian Toda theories contain spin-one
fields. Let us perform our calculations on the Bershadsky algebra already
considered in the previous sections (see in particular Eq. (\ref{eq:4.4})).

First, by the following shift on $T$,
\be
\bar{T} = T - \sm{1}{3c} J^2
\ee
one gets the P.B.:
\bea
\{\bar{T}(z) , J(w) \} &=&0 \nonumber \\
\{ \bar{T}(z) , W_\pm (w) \} &=& - \smt W_\pm (w) \delta ' (z-w)
+ (\cd W_\pm)(w) \delta(z-w) \nonumber \\
\{ W_+(z), W_-(w) \} &=& (\bar{T} - c\cd^2 ) (w) \delta (z-w)
\ena
while  $\bar{T}$ satisfies the usual Virasoro P.B.:
\be
\{ \bar{T}(z), \bar{T}(w) \} = -2\bar{T}(w) \ \delta ' (z-w)
+ \prt \bar{T}(w) \delta (z-w) + \sm{c}{2}
\delta ''' (z-w)
\ee

In the above equations, one has used the covariant derivative $\cd$ such that
\be
\cd W_\pm = (\prt \mp \sm{1}{c}J ) W_\pm
\label{eq:6.4}
\ee
while the $D^2$ showing up in the r.h.s. of $\{ W_+,W_- \}$ is relative to $w$.
The appearance of a covariant derivative may open new perspectives in the field
of integrable models. It is here particularly convenient in order
to construct the
commutant of $J$. Indeed the set of fields commuting with $J$ is generated by
the stress energy tensor $\bar{T}$ and the bilinear products:
\be
W^{(p,q)} = ({\cd}^p W_+)({\cd}^qW_-)
\ee
with $p,q$ non negative integers.

Actually, the fields $W^{(p,q)}$ and $\bar{T}$
are the building blocks from which
one can construct an infinite tower of primary fields of spin 3,4,...
\bea
W_3 &=& W_+ W_- \nonumber \\
W_4 &=& W_+ {\cd}W_- - W_- \cd W_+ \nonumber \\
&& \vdots\nonumber \\
W_{3+n} &=& W_+\cd^nW_- -(\cd^nW_+)W_- +\dots \mb{for}n>2
\ena
these fields being created by the P.B. of fields of lower conformal spin, for
ex.:
\be
\{ W_3(z), W_3(w) \} = 2W_4(w) \delta' (z-w) - \prt W_4 (w) \delta(z-w)
\ee
and so on.

At this point, one may say that by looking at the commutant of the spin one
generator $J$ in the Bershadsky $W$ algebra, one has obtained a polynomial non
linear \underline{$W_\infty$ realization}.

But the primary fields $W_{3+n}$ with $n \geq 2$ are  not independent, and can
be expressed as \underline{rational} -and not polynomials- functions of $T,
W_3, W_4$: for example $W_5$ can be written in terms of $W_3$ and $W_4$ as
follows:
\be
W_5 = \frac{1}{4W_3} \left[ 7 \left(W^2_4 - (\prt W_3)^2 \right)
+ 6W_3 (\prt^2 W_3) + \bar{T}W_3 ) \right]
\ee

Therefore, the commutant of $J$ exhibits a new structure with respect to the
standard $W$ algebras, which can be seen either as a rational finitely
generated $W$ algebra or as a polynomial non linear $W_\infty$ realization.

The above example is the simplest one exhibiting such a structure. Of course a
general approach with a non Abelian $W_1$ part can be performed (see
\cite{Wrat}).

\subsection{Supersymmetric extension}

\indent

The supersymmetric extension of this problem can be considered in an analogous
way. Again, let us illustrate the method on an example, the $N=3$
superconformal algebra ${\cs}{\cc} (N=3)$ generated by a spin 2 generator
$T(z)$, 3 spin $\smt$
components $G^a_\smtt\ (a=1,2,3)$, 3 spin 1 elements $J^a(z)$, constituting an
$Sl(2)$ Kac-Moody algebra and a spin $\shalf$ fermion $\psi(z)$. The C.R. in
the classical case can be deduced from the formulas (15) of \cite{god}, in
which we identify the O.P.E. with the P.B. and the singular terms
$\frac{1}{(z-w)^k}$ with $(-1)^{k-1}\frac{1}{(k-1)!} \delta^{(k-1)} (z-w)$.
After defining:
\be
\begin{array}{lll}
G^\pm(z) = \frac{1}{\sqrt{2}} (G^1 \pm iG^2) (z) & \mbox{and} & J^\pm(z) =
\frac{1}{\sqrt{2}} (J^1 \pm iJ^2)(z) \\
G^0(z) = G^3(z) && J^0(z) = J^3(z)
\end{array}
\ee
we will adopt the superfield formalism (cf. \cite{supertoda}) and define:
\be
\begin{array}{ll}
{\ct}(z)  = \shalf \ G^0(z) + \theta T(z) & \mb{of superspin} \smt \\
{\cj}^\pm(z)  = \pm J^\pm(z) + \theta G^\pm(z) & \mb{of superspin} 1 \\
\Phi(z) = \psi(z) + \theta J^0(z) & \mb{of superspin} \shalf
\end{array}
\ee
using the supervariable notations:
\be
Z=(z,\theta), \ \ W=(w,\eta) \ \ \mbox{and} \ \ Z-W=z-w-\theta \eta
\ee
then the P.B. can be "compactly" written as (keeping in mind from above that:
$\frac{\theta -\eta}{Z-W} = (\theta - \eta) \delta (Z-W) \doteq \delta (Z-W)$
and so on for their derivatives, and the O.P.E. being in place of the P.B.):
\be
{\ct}(Z) \cdot \Theta_s (W) = s \frac{\theta-\eta}{(Z-W)^2} \Theta_s(W) +
\shalf \ \frac{D \theta_s(W)}{Z-W} + \frac{\theta - \eta}{Z-W} \ \prt
\Theta_s(W)+ \dots
\ee
if $\Theta_s(W)$ denotes the superspin ${\cj}^\pm (W)$ or $\Phi(W)$ of
superspin $s=1$ or $\shalf$, and as usual: $D=\prt_\eta + \eta \prt_w$
\bea
{\ct}(Z) {\ct}(W) &=& \smt \ \frac{\theta - \eta}{(Z-W)^2} {\ct}(W) + \shalf
\ \frac{D {\ct}(W)}{Z-W} + \frac{\theta -\eta}{Z-W} \prt{\ct}(W) +
\frac{c/6}{(Z-W)} + \dots \nonumber \\
\Phi(Z) {\cj}^\pm (W) &=& \pm \frac{\theta - \eta}{Z-W} {\cj}^\pm (W) + \dots
\nonumber \\
\Phi(Z) \Phi(W) &=& \frac{c/3}{Z-W} + \dots \nonumber \\
{\cj}^+ (Z) {\cj}^-(W) &=& - \frac{\theta-\eta}{(Z-W)^2} \Phi(W) -
\frac{1}{Z-W}
D\Phi(W) - \frac{\theta-\eta}{Z-W} \prt \Phi  \nonumber \\
&& -2 \frac{\theta-\eta}{Z-W} \ct(W) - \frac{c/3}{(Z-W)^2} + \dots
\ena

We wish to factorize out the superspin $\shalf$ superfield $\Phi(Z)$. As in the
nonsupersymmetric case, we can operate a shift on ${\ct}(Z)$
\be
{\ct}_0(Z) = {\ct}(Z) - \frac{3}{2c} \Phi(Z) D \Phi (Z)
\ee
such that:
\be
{\ct}_0(Z) \cdot \Phi(W) =0
\ee

We can expect the covariant derivative of Eq.(\ref{eq:6.4}) to become:
\be
{\cd} = D - \frac{3q}{c} \Phi
\ee
if $q$ is the super $U(1)$ charge carried by the primary superfield, i.e.:
\be
{\cd} {\cj}^\pm = (D \mp \sm{3}{c} \Phi) {\cj}^\pm
\ee

Now the spin 2 superfield $W_2(Z) = {\cj}^+ (Z) \cdot {\cj}^-(e)$ is a primary
superfield under ${\ct}_0(Z)$ in the commutant of $\Phi(Z)$. The properties
above obtained with $W$ algebras generalize here with $W$ superalgebras.
Computing for example the P.B. of $W_2$ with itself one gets:
\bea
W_2(Z) W_2(W) &=& - \sm{c}{3} \left( \frac{2W_2(W)}{(Z-W)^2} + \frac{\prt
W_2(W)}{Z-W} + \frac{\theta-\eta}{(Z-W)^2} DW_2(W) \right. \nonumber \\
&& \left. + \sm{3}{5} \ \frac{\theta - \eta}{Z-W} D \prt W_2(W) \right) -
\sm{36}{5} \ \frac{\theta - \eta}{Z-W} ({\ct}_0 \cdot W_2) (W) \nonumber \\
&& + \sm{c}{3} \ \frac{\theta -\eta}{Z-W} W_{7/2} (W) + \dots
\ena
where $W_{7/2}(W)$ is the (new!) $7/2$ superspin primary superfield defined as:
\be
W_{7/2} = {\cj}^+ {\cd}^3 {\cj}^- + {\cj}^- {\cd}^3 {\cj}^+ - \sm{3}{5} D \prt
W_2 - \sm{48}{5c} \ {\ct}_0 \cdot W_2
\ee

\subsection{Spin 1/2 versus spin 1 fields}

\indent

The superalgebra ${\cs \cc} (N=3)$ was the first example considered by the
authors of \cite{god} to illustrate their result about
the factorization of the
spin $1/2$ part in a superconformal field theory, more precisely that a
meromorphic field theory can be decomposed into the tensor product
of a spin $1/2$ part and
a conformal field theory without spin $1/2$ field. We would like to
stress that
this property can easily be proved, at least at the classical level, by the use
of finite gauge transformations already introduced in the previous section (see
Eq.(\ref{eq:5.8}). Indeed, leaving to the reader the general proof (which will
also be found in \cite{prep}) let us stay with the ${\cs \cc}(N=3)$ algebra
and perform on its generators $X(w)$ the transformation:
\be
X(w) \rightarrow \hat{X}(w) = X(w) + \int dz\ \al(z) \psi (z) . X(w)
+ 0
\ee
where $\psi(z)$ is the fermion field (we do not use any more the superfield
formalism, since we wish to only factorize the $\psi(z)$ fermion and not the
superspin $1/2$ field).

Owing to the OPE relation:
\be
\psi(z) \cdot \psi(w) = \frac{c/3}{z-w}
\ee
one directly gets, imposing the "gauge fixing":
\be
\al(w) = - \psi(w)
\ee
the transformed fields:
\be
\hat{\psi} = 0 \ \ ;\ \ \hat{T} = T - \shalf \psi \prt \psi \ \
;\ \ \hat{G}^a = G^a - T^a\psi\ \; \ \ \hat{J^a} = J^a\ \
\mb{with}a=1,2,3
\ee
In accordance with the results of \cite{god}, the O.P.E. among the
transformed fields are identical, except for the central charge to the ones
relative to the non transformed fields, and as expected such that:
\be
\hat{T} \cdot \psi = \hat{G}^a \cdot \psi = \hat{J}^a \cdot \psi =0
\ee

Note that this gauge transformation can also be done with spin 1/2
bosons, and leads to the same conclusion \cite{prep}. It has also be shown
that
the action of such a super-Toda model can be rewritten as the sum of
two terms, one relative to the spin 1/2 part and the other to the
factorized $W$ part \cite{rag}.

It is natural to wonder what happens if, instead of performing a gauge
transformation associated with a $1/2$ fermion, one involves a
spin 1 field. Let
us take once more as an example the Bershadsky algebra (see
Eq.(\ref{eq:4.4})):
its (simple) Kac Moody generator $J(z)$ satisfies:
\be
J(z) \cdot J(w) = \frac{3/2 c}{(z-w)^2}
\ee
In order to obtain $\hat{J} =0$ in the transformation:
\be
J(w) \rightarrow \hat{J}(w) = J(w) + \int dz\ \al(z) J(z) \cdot J(w) + \dots
\ee

We would have to impose $\al$ such that
\be
\prt \al (w) = J(w)
\ee

The pathology created by this relation appears in different places. In
particular, one would get:
\be
\al(z) \cdot W^\pm_{\smtt} (w) = \pm \smt W^\pm_\smtt (w) \ln (z-w)
\ee
and some trouble to compute, from:
\be
\hat{W}^\pm_\smtt (w) = e^{\pm \smtt \al(w)} W^\pm_\smtt (w)
\ee
the quantity:
\be
\hat{W}^+_\smtt(z) \cdot \hat{W}^-_\smtt (w)
\ee
Thus, gauge transformations relative to spin $1/2$ fields allow to recover the
result of Ref \cite{god}, namely the property that spin $1/2$ fermions can be
eliminated in a super $W$ algebra, but such a technics does not appear suitable
for the factorization of spin 1 fields, as could be expected from the results
presented in the first part of this section.

Note that the above discussion has to be compared with the
factorization at quantum level, of spin 1/2 and 1 fields considered in
\cite{dec}: the projection used there appears as a quantum
version of our gauge transformation.

\vs{5}

\noindent
{\Large{\bf Acknowledgements}}

\indent

It is a pleasure to thank F. TOPPAN for discussions. Paul
SORBA is endebted to the organizers of the Conference for the pleasant and
warming atmosphere during the meeting.

\end{document}